\documentclass[12pt, showkeys]{JHEP3}
\usepackage{epsfig}
\usepackage{amsmath}
\def\beq{\begin{eqnarray}}
\def\eeq{\end{eqnarray}}

\newcommand{\de}{\partial}
\newcommand{\be}{\begin{equation}}
\newcommand{\ba}{\begin{eqnarray}}
\newcommand{\ea}{\end{eqnarray}}
\newcommand{\ee}{\end{equation}}

\renewcommand{\beq}{\begin{equation}}
\renewcommand{\eeq}{\end{equation}}
\newcommand{\beqa}{\begin{eqnarray}}
\newcommand{\eeqa}{\end{eqnarray}}
\newcommand{\CR}{\nonumber \\}

\newcommand{\bra}[1]{\left\langle\, #1\,\right|}
\newcommand{\ket}[1]{\left|\, #1\,\right\rangle}

\newcommand{\ra}{\rightarrow}

\newcommand{\bR}{{\bf R}}
\newcommand{\bZ}{{\bf Z}}
\newcommand{\Dslash}{D\!\!\!\!\slash\,}

\newcommand{\dz}{\mbox{D}0{\rm -}\overline{\mbox{D}0}}

\def\Tr{\mathop{\rm Tr}\nolimits}

\def\mat#1{\matt[#1]}
\def\matt[#1,#2,#3,#4]{\left(%
\begin{array}{cc} #1 & #2 \\ #3 & #4 \end{array} \right)}

\title{
Tachyon Condensation on Torus 
and T-duality 
}
\author{
Seiji Terashima\\
{\it Yukawa Institute for Theoretical Physics, 
Kyoto University, \\
Kyoto 606-8502, Japan}\\
}

\abstract{
We find an exact solution, with a nonzero net D-brane charge,
in the boundary string field theory 
of brane-anti-brane pairs on a torus.
We explicitly take the T-dual of this configuration.
The Nahm-transformation of the instantons 
is derived from the tachyon condensation.
}

\preprint{
{\normalsize YITP-08-44}
}

\begin{document}

\section{Introduction}

The open sting tachyon condensation on unstable D-branes
have been intensively investigated 
in the past decade \cite{Sen} \cite{Senreview}.
Among them, the exact solutions of the tachyon condensation,
were found in the boundary string field theory\footnote{
Recently, the exact solution in the Witten's cubic string 
field theory for the bosonic string was found in \cite{Schnabl}.
it may represent a closed string vacuum.} 
\cite{KMM1, GeSh} \cite{KMM2, KrLa, TaTeUe}
or in the boundary state \cite{AsSuTe3} on $\bR^r$.\footnote{
Recently, the boundary string field theory 
was reconstructed via the boundary state 
\cite{Teraguchi, IsTe}.}
Those solutions include the topologically non-trivial 
solutions, for example, the kink or the vortex,
which represent lower dimensional D-branes,
with codimension one and two, respectively.
This constructions of lower dimensional D-branes 
from unstable D-branes by the tachyon condensation were
known as the decent relations \cite{Sen}.
On the other hand, 
we can construct higher dimensional D-branes 
from lower dimensional unstable D-branes,
like the matrix models, 
by the tachyon condensation on $\bR^r$ \cite{Te, AsSuTe1, AsSuTe2},
which were known as the ascent relations.

Since $\bR^r$ is topologically trivial and non-compact,
there is no winding modes and
the solutions in the boundary string field theory 
can have a trivial bundle.
Since the torus is simplest non trivial compact manifold,
the study of the tachyon condensation on torus 
will be interesting.\footnote{
On a torus with the self dual radius, 
an exact solution of the tachyon condensation of 
a $\mbox{D}{\rm -}\overline{\mbox{D}}$-brane pairs
was given in \cite{Sen, MaSe} by the marginal deformations
(the ``tachyon'' is massless).
This solution corresponds to the 
lower dimensional 
$\mbox{D}{\rm -}\overline{\mbox{D}}$-brane systems. 
By the marginal deformation, 
we always has a $\mbox{D}{\rm -}\overline{\mbox{D}}$-brane
or non BPS D-branes which do not have net D-brane charges because of
the charge conservation. In this paper we will study 
the soliton with a net D-brane charge.
}
However, it will be very difficult 
to find an exact solution of a soliton on
a $\mbox{D}{\rm-}\overline{\mbox{D}}$-brane pair
on a torus. 
To be explicit, let us consider a vortex soliton of the tachyon of
a $\mbox{D}2{\rm -}\overline{\mbox{D}2}$-brane pair
on $T^2$. This soliton will represent a D0-brane. 
Since the torus is an orbifold of $\bR^2$,
it seems easy to construct such soliton, 
however, the orbifolding of the solution is not straightforward.
Actually, on $\bR^2$ the D0-brane solution is represented
by a following non-periodic configuration:
\begin{eqnarray}
 && T= u (\xi^1+ i \xi^2), \;\; (u \rightarrow \infty), \CR
 && A^{(1)}_\mu=A^{(2)}_\mu=0, \;\; (\mu=1,2),
\label{vortex}
\end{eqnarray}
in the boundary string field theory.
Here $T$ is the tachyon and 
$A_\mu^{(1)}$ and $A_\mu^{(2)}$ are the gauge fields
on the D2-brane and anti-D2-brane, respectively.
This is the exact solution of the equations of motion
in the $u \ra \infty$ limit. 
Obviously, it is difficult to extend this solution (\ref{vortex})
on $\bR^2$ 
to a solution on $T^2=\bR^2/\bZ^2$ because of the 
non periodicity of (\ref{vortex}).\footnote{
If we take the bundle on the D2-brane as (\ref{bdld2})
and the trivial bundle on the anti-D2-brane,
a general tachyon field would be written as
\begin{equation}
 T(\xi^1,\xi^2)=u \left(
\sum_{n \in \bZ} 
H \left( n+\frac{\xi^1}{2 \pi L_1} \right) e^{i \frac{\xi^2}{L_2} n}
\right) G(x^2,x^2),
\end{equation}
where $G(\xi^1,\xi^2)$ is a periodic function of $\xi^\mu$.
}
Moreover, the gauge fields can not be trivial on $T^2$
because it is a compact space.
It will be interesting, but, highly non-trivial to
construct a vortex solution 
(${\cal D}_\mu T \sim 0$) with non-zero $A_\mu^{(i)}$.

In this paper, we consider the tachyon condensation of 
$\mbox{D}{\rm -}\overline{\mbox{D}}$-brane pairs
on a torus and
find exact solutions, which have a net D-brane charge,
in the boundary string field theory or 
the boundary state formalism.
Our construction uses infinitely many 
$\mbox{D}{\rm -}\overline{\mbox{D}}$-brane pairs
instead of a pair.
By the T-dual transformation,
we can change the dimensions of 
the $\mbox{D}{\rm -}\overline{\mbox{D}}$-brane pairs
and we will see that the 
soliton on the $\mbox{D}0{\rm -}\overline{\mbox{D}}$-brane pairs
which represents a D$(2p)$-brane is 
the simplest one.

As an application of the soliton solution,
we can consider the Nahm-transformation \cite{BrBa, Sc}
which maps an anti-self dual gauge field 
(instanton) of U(N) with the instanton number $k$ on 
four dimensional torus $T^4$
to an anti-self dual configuration of U(k) 
with the instanton number $N$ on 
on a dual torus $\tilde{T}^4$.
In string theory, 
the bound state of 
$N$ D$(p+4)$-branes and $k$ D$p$-branes on $T^4$
is given by the $U(N)$ gauge instanton.
If we introduce a probe D$(p-4)$-brane
and consider the low energy limit on it,
the Nahm transformation was 
interpreted as the T-dual transformation \cite{Ho},
generalizing the ADHM(N) cases \cite{Do2} \cite{Di}.
Recently, 
the ADHM transformation was given in
a D-brane setup
without a probe D-brane and nor 
a low energy limit \cite{HaTe3, HaTe4}
by the method using the tachyon condensation \cite{Te1, Te2}.
For a bound state
of two different D-branes,
say the D$(p+q)$-branes and the D$p$-branes,
this method gives the equivalence between 
the descriptions using the D$(p+q)$-branes and 
using the D$p$-branes.
This method can be applied to 
$T^4$ case and we will see that
the Nahm transformation is naturally interpreted
as this equivalence (plus the T-dual transformation).
It is worth noticing that
this equivalence is exact in $\alpha'$,
therefore, the $N$ D4-brane with the $k$ instanton on $T^4$ is 
equivalent to $k$ D4-branes with 
the $N$ instanton on the T-dual torus $\tilde{T}^4$,
which has a sub-stringy size if the size of $T^4$ is much bigger
than string scale.\footnote{
We assume we can employ off-shell boundary states, which are naive extensions of 
the boundary state, as in \cite{HaTe3, HaTe4}. 
They have possibility of suffering from divergences 
when away from on-shell background fields. 
However, the off-shell boundary states have a natural interpretation 
in consistency with the boundary string field theories.
Furthermore, 
our main concern is the on-shell configurations
although finding those are not discussed in this paper.
Actually, the instanton configurations on torus
is expected to be on-shell for all order in $\alpha'$ as discussed 
in \cite{HaTe3,HaTe4}.}

This paper is organized as follows.
In section 2, 
we review how to obtain the T-dual picture
of the D0-branes on torus, according to \cite{Ta}.
In section 3, 
we give an exact solution
in the boundary string field theory 
of brane-anti-brane pairs on torus.
We take the T-dual of this configuration.
The Nahm-transformation of the instantons 
is derived from the tachyon condensation.
We conclude with some discussions in section 4.

\section{D0-branes on Torus and T-dual}

In this section, we will review how to
describe D0-branes in type II superstring theory
on a (rectangular) torus as an orbifold $T^r=\bR^r/\bZ^r$,
whose periodic coordinates $0 \leq x^\mu < 2 \pi R_\mu$,  
according to \cite{Ta,DoMo} and how to take the T-dual of the D0-branes
\cite{Ta}.

We consider only the scalars corresponding to 
the locations of the D0-branes in the torus, $X^\mu(t)$, where
$\mu=1, \ldots, r$
although there are many fields on the D0-branes.
In this paper, 
the time $t$ is always fixed, and thus abbreviated below.
If we consider the time-independent $X^\mu$,
it can be considered as the static configuration in the  $A_0=0$ gauge.     

Since $T^r=\bR^r/\bZ^r$, the $N$ D-branes on tours will be
equivalent to the $N \times \infty$ D-branes on $\bR^r$ 
whose coordinates $X^\mu$ 
will be operator valued Hermite $N \times N$ matrices.
Here we regard an $\infty \times \infty$ matrix as 
an operator.
   
By the orbifolding, we need the following identification with translation 
operators $U'_\nu$ along $x^\nu$
which should be operator valued $N \times N$ unitary matrices:
\beq
U'_\nu X^\mu {U'_\nu}^{-1} = 
\Omega'_\nu 
\left( X^\mu + \delta_{\mu \nu} 2 \pi R_\nu  1_{N} \right) 
{\Omega'}_\nu^{-1},
\label{trans0}
\eeq
where 
$\Omega'_\nu$ is an $N \times N$ unitary matrix, i.e. a gauge
transformation
of the $N$ D-branes.
Throughout this paper, we take a convention that 
an index $\nu$ is not summed over except explicitly 
indicated by $\sum_\nu$.
We will define $U_\nu = {\Omega'}_\nu^{-1} U'_\nu$,
then 
\beq
U_\nu X^\mu {U_\nu}^{-1} = 
X^\mu + \delta_{\mu \nu} 2 \pi R_\nu  1_{N}.
\label{trans}
\eeq
A representation of (\ref{trans}) is
\beqa
X^\mu &=& 2 \pi \alpha' 
\left( i \frac{\de}{\de \xi^\mu} + \tilde{A}_\mu(\xi) \right) \CR
U_\nu &=&  e^{i\frac{\xi^\nu}{L_\nu}},
\label{rep1}
\eeqa 
where $\tilde{A}_\mu(\xi)$  is an $N \times N$ matrix and  
\beq
L_\nu=\frac{\alpha'}{R_\nu}.
\eeq
Here $\xi_\nu$ is the periodic coordinate of 
the T-dual torus $\tilde{T}^r$ and 
$0 \leq \xi_\nu < 2 \pi L_\nu$
and $\tilde{A}_\mu$ is the gauge field of the $N$ D$r$-branes 
on $\tilde{T}^r$.\footnote{
The gauge transformation of the $N$ D0-branes 
should not change the (\ref{trans}).
Thus the transformation is generated by 
a $N \times N$ unitary matrix 
$U_{N \times N} (\xi, \frac{\de}{\de \xi})$ which 
commutes with $U_\nu$.
This is actually a gauge transformation of 
the $N$ D$r$-branes, i.e. a unitary matrix 
$U_{N \times N}(\xi)$.
}
Note that this implies that
the gauge field $\tilde{A}_\mu(\xi)$ is a connection 
on the $\tilde{T}^{r}$, whose component is 
not necessary a periodic function of $\xi$.
If the bundle of the $N$ D$r$-branes on $\tilde{T}^r$
with the gauge field $\tilde{A}_\mu(\xi)$ is trivial, i.e. 
$\tilde{A}_\mu(\xi)$ is periodic,
the base of the Hilbert space is spanned by
\beq
e^{-i \sum_{\nu=1}^r \left( \frac{\xi^\nu n_\nu}{L_\nu} \right)} v_N,
\eeq
where $n_\nu \in \bZ$ and $v_N$ is a base of a $N$-vector.
If the bundle of D$r$-branes is non-trivial,
the base of the Hilbert space 
will be the sections of the bundle on the dual torus $\tilde{T}^r$.

Finally, let us consider a bound state 
of a D0-brane and $m$ D$r$-branes on the torus $T^r$.
First, we sketch how to construct a D$r$-brane within $m$ D$0$-branes
on the T-dual torus $\tilde{T}^r$. 
We will consider $r=2$ case as an example.
The bound state of the D2-brane and the D0-branes 
on $\tilde{T}^2$
will be given by 
\begin{eqnarray}
 && \tilde{A}_1=0, \,\, \tilde{A}_2= \tilde{F} \xi^1, \,\,\,\,\, \CR
 && \tilde{F}=\frac{m}{2 \pi L_1 L_2} ,
\end{eqnarray}
where $m$ is an integer which is the D0-brane charge.
The transition function (or the gauge transformation) 
between the different patches is given by
$\tilde{A}_\mu(\xi^1+2 \pi L_1,\xi^2)
=\tilde{\Omega}_1 \tilde{A}_\mu \tilde{\Omega}_1^{-1} -i (\de_\mu \tilde{\Omega}_1) \tilde{\Omega}_1^{-1}$
and $\tilde{A}_\mu(\xi^1,\xi^2+2 \pi L_2)
=\tilde{\Omega}_2 \tilde{A}_\mu \tilde{\Omega}_2^{-1} -i (\de_\mu \tilde{\Omega}_2) \tilde{\Omega}_2^{-1}$
where
\begin{equation}
 \tilde{\Omega}_1=e^{ i \frac{\xi^2}{L_2}}, \;\;\; \tilde{\Omega}_2=1. 
\label{bdld2}
\end{equation}
Note that this is the exact solution.
Here an exact solution means that
a solution of the equations of motions of the D2-brane 
(string field theory) action
including all order in the $\alpha'$ expansions, but leading order 
in the string coupling $g_s$.
Then, from the T-dual map (\ref{rep1}), 
we can read the D0-brane configuration $X^\mu$
of the bound state of the D0-brane and the $m$ D$r$-branes
on $T^r$.

\section{$\mbox{D}0{\rm -}\overline{\mbox{D}0}$ pairs on Torus}

In this section, 
we will construct 
a solution which is equivalent to $M$ D($2p$)-branes
in the boundary string field theory 
of infinitely many $\dz$-brane pairs
on a torus $T^{2p}$.

First, we consider the infinitely many $\dz$-brane pairs on $\bR^{2p}$.
The solution which is equivalent to the $M$ D($2p$)-branes 
$\bR^{2p}$ with gauge field $A_\mu(x)$ is 
\beq
\mat{0,T,T^\dagger,0} =\lim_{u \ra \infty} u \;  
\Gamma^\mu \otimes (1_{M \times M} \otimes 
\hat p_\mu- A_\mu(\hat x) ), 
\;\;\;\;\; X^\mu=
1_{2^{p} \times 2^{p}} \otimes 1_{M \times M} 
\otimes \hat x^\mu, 
\label{solt}
\eeq
where $X^\mu$ is the transverse scalars of D0-branes and
the $T$ is the tachyon which acts on the D0-branes and 
$T^\dagger$ acts on the anti-D0-branes, which correspond
to the anti-chiral spinors.
Here we have set that
the anti-D0-branes has the transverse scalars with the same v.e.v 
as the D0-branes.
The operators $\hat{x}^\mu, \hat{p}_\mu $ satisfy
$[\hat{x}^\mu, \hat{p}_\nu ]= i \delta_{\mu,\nu}$ and
$\Gamma^\mu$ is the 
Dirac gamma matrix of $SO(2p)$ which satisfies
$\Gamma \equiv i^{-p} \Gamma^1 \Gamma^2 \cdots \Gamma^{2p}
=\mat{1_{2^{p-1} \times 2^{p-1}},0, 
0,-1_{2^{p-1} \times 2^{p-1}}}$. 
Note that $T$ and $X^\mu$ act on 
the Dirac spinors which transformed as 
a fundamental representation 
of the $U(M)$ gauge symmetry on the manifold spanned 
by $M$ D$(2p)$-branes.
Thus the (\ref{solt}) can be written as
\beq
\mat{0,T,T^\dagger,0}=\lim_{u \ra \infty} u \;  \Dslash,
\;\;\;\;\; X^\mu=x^\mu. 
\eeq

Using this configuration 
we can construct the $M$ D($2p$)-branes with gauge
field $A_\mu(x)$
on the torus $T^{2p}$, 
which is spanned by $0 \leq x^\mu \leq 2 \pi R_\mu$,
by the orbifolding of $\bR^{2p}$.
Here we assume that 
$A_\mu(x)$ 
satisfies
\beq
A_\rho(x_\mu+\delta^{\mu \nu} 2 \pi R_\nu ) 
=\Omega_\nu A_\rho(x) \Omega_\nu^{-1}
- i \Omega_\nu \de_\rho \Omega_\nu^{-1},
\eeq 
where $\Omega_\nu$ is a transition function (or gauge transformation)
on the torus.
Thus the $A_\mu$ is a gauge field on $\bR^{2p}$
which is extended from the gauge field on the torus,
i.e. a pull back connection of the map from
$\bR^{2p}$ to $T^{2p}=\bR^{2p}/\bZ_{2p}$.
The constraint for the tachyon $T$ by the orbifolding 
may be same as the constraint for transverse coordinates. 
Thus we require that
\beqa
U_\nu X^\mu U_\nu^{-1} &=& 
X^\mu + \delta_{\mu \nu} 2 \pi R_\nu  1_{N} 
, \CR
U_\nu T U_\nu^{-1} &=& 
T,
\label{const1}
\eeqa
where we take same $\Omega_\nu$ in (\ref{trans0}) for the D0-branes and 
the anti-D0-branes.
Then the configuration (\ref{solt}) is consistent 
with the constraint (\ref{const1}) of the orbifolding 
if we take
\beq
U_\nu=  \Omega_\nu \, e^{2 \pi i \hat p_\nu R_\nu}.
\label{uorb}
\eeq 
This is obvious if we notice 
that the $\dz$-branes given by the configuration (\ref{solt}) 
uniformly distributed in $\bR^{2p}$ and
the unit shift ($\ref{uorb}$) is a symmetry.
Note that 
\beq
U_\rho U_\mu=U_\mu U_\rho,
\eeq
which is from the fundamental 
property of the transition functions.

Therefore, 
the configuration (\ref{solt}) with the orbifolding operator
(\ref{uorb}) is a consistent configuration of
the infinitely many $\dz$-brane pairs
on a torus $T^{2p}$
which is equivalent to $M$ D($2p$)-branes
with the gauge field $A_\mu(x)$.
Since the orbifolding will consistently truncate the equations 
of motion or the (on-shell) boundary state, 
(\ref{solt}) with the orbifolding by the generator (\ref{uorb})
will be an exact solution on the torus if we set $A_\mu=0$ 
or, for example, 
an anti self-dual configuration for $p=2$.\footnote{
It is desirable and interesting to study the 
solution on the torus
in the boundary state formalism.}

\subsection{
Nahm Transformation and
Tachyon condensation on 
$\mbox{D}0{\rm -}\overline{\mbox{D}0}$ pairs 
}

If we consider $M$ D(2p)-branes with
a nontrivial gauge bundle on the torus,
it is 
the bound state of $M$ D(2p)-branes and the lower 
dimensional D-branes, for example D0-branes.
In this case,  following  \cite{Te1} (see also \cite{El}) 
we can find a configuration of D0-branes which 
is equivalent to the bound state.
We will apply this to the solution on the torus
and see that the Nahm transformation naturally appears.

In \cite{Te1},
we first take a configuration of 
$\mbox{D}0{\rm -}\overline{\mbox{D}0}$ pairs which represents
the $M$ D(2p)-branes by the tachyon condensation.
Then the tachyon is diagonalized by the gauge transformation
and then only D0-branes which corresponds to 
zero modes remain after the tachyon condensation,
namely the $u \ra \infty$ limit.
Then we see that the remaining D0-branes have the transverse scalars
or the matrix coordinate $\bar{X}^\mu$ is 
given by just a truncation of the Chan-Patton-Hilbert space
to those composed by the the zero modes only:
\beq
\left(\bar{X}^\mu  \right)^{i}_{\;\;\;\; j}
=\bra{i} X^\mu \ket{j},
\eeq
where
$\ket{i}$ is a zero mode of the tachyon.
This gives the D0-brane picture of the boundary state.

For the $M$ D(2p)-branes with
a nontrivial gauge bundle on the torus,
the tachyon,
\beq
\mat{0,T,T^\dagger,0}=\lim_{u \ra \infty} u \; \Dslash,
\label{diracop}
\eeq
acts on $\Psi(x)$ which is a spinor on $\bR^{2p}$.
Now we decompose a spinor on $\bR^{2p}$ into
a spinor on torus and a plain wave 
like the Bloch wave function:
\beq
\Psi_{\xi}(x)= e^{i \frac{1}{2 \pi \alpha'} \xi_\mu  x^\mu  }
\psi_\xi (x)
\label{decomp}
\eeq
where $\psi_\xi(x)$ is
\beq
U_\nu \psi_\xi(x) =  \psi_\xi(x),
\eeq
which means that $\psi_\xi(x)$ is a section of the 
spinor bundle on $T^{2p}$
and $0 \leq \xi_\mu < 2 \pi L_\mu$.
Indeed, $\Psi_\xi$ is the eigen state of 
the unitary operator $U_\nu$:
\beq
U_\nu \Psi_\xi (x) = e^{i \frac{\xi_\nu R_\nu}{\alpha'}}
\Psi_\xi (x).
\eeq
Thus any spinor $\Psi(x)$ on $\bR^{2p}$ can be
written as
\beqa
\Psi(x) 
 &=& \int_0^{\frac{2 \pi \alpha'}{R_1}} d \xi_1 
\int_0^{\frac{2 \pi \alpha'}{R_2}} d \xi_2 \cdots
\int_0^{\frac{2 \pi \alpha'}{R_{2p}}} d \xi_{2p} 
e^{i \frac{1}{2 \pi \alpha'} \xi_\mu  x^\mu  } \;
\psi_\xi(x),
\eeqa
because any eigen state of $U_\nu$ can be 
written as (\ref{decomp}).
Using this, we have
\beq
\Dslash \Psi(x)= \int_0^{\frac{2 \pi \alpha'}{R_1}} d \xi_1 
\int_0^{\frac{2 \pi \alpha'}{R_2}} d \xi_2 \cdots
\int_0^{\frac{2 \pi \alpha'}{R_{2p}}} d \xi_{2p} 
e^{i \frac{1}{2 \pi \alpha'} \xi_\mu  x^\mu  } \;
\Dslash_\xi \psi_\xi(x),
\eeq
where 
\beq
\Dslash_\xi =
\Gamma^\mu \left( \hat p_\mu - A_\mu(\hat x) 
+\frac{\xi_\mu}{2 \pi \alpha'} \right).
\eeq
Then, the zero modes of the tachyon, $\Dslash \Psi(x)=0$, is 
given by 
\beq
\Psi_{\xi}^{i}(x)= e^{i \frac{1}{2 \pi \alpha'} \xi_\mu  x^\mu  }
\psi_\xi^i (x)
\eeq
where $\psi_\xi^i(x)$ is a zero mode of $\Dslash_\xi$, i.e. it satisfies
\beq
\Dslash_\xi \psi_\xi^i(x)=0,
\label{dpsi}
\eeq
$(i=1,\cdots,m)$ and $m$ is the number of the zero modes
of $\Dslash_\xi$. \footnote{
Here we assume that $m>0$ and the all zero modes 
have positive chirality, i.e. $\Gamma \psi_\xi^i(x)=\psi_\xi^i(x)$, 
which means that only the $m$ D0-branes are remained
and all anti-D0-branes
disappear after the tachyon condensation.
However, this assumptions is not essential, even for 
cases with zero modes of both chiralities, as
discussed in \cite{HaTe3, HaTe4}.}
From the index theorem \cite{index}, 
we know that $m$ does not depend on $\xi$.
For $p=2$,
$m$ is the instanton number.
The zero modes are labeled by $\xi$ and $i$.
We will see that 
the discrete eigen values of 
$\frac{\de}{\de \xi^\mu}$
parameterize 
mirror images of D0-branes by the 
$\bZ^{2p}$ orbifolding.

We normalize the zero modes as
\beq
\frac{1}{2 \pi} \int_{\bR^{2p}} d^{2p} x \, \Psi_{\xi}^{i}(x)^\dagger \,
\Psi_{\xi'}^{j}(x)
= \delta (\xi-\xi') \delta_{ij},
\eeq
which is equivalent to 
\beq
\int_{T^{2p}} d^{2p} x' \, \psi_{\xi}^{i}(x')^\dagger \,
\psi_{\xi}^{j}(x')
=\delta_{ij},
\eeq
where $0 \leq {x'}^\nu \leq 2 \pi R_\nu$.
(Because of the Euclidean nature, $\Psi^\dagger \Psi$
is the $SO(2p)$ invariant.)
This can been seen from
\beqa
&& \frac{1}{2 \pi} \int_{\bR^{2p}} d^{2p} x \, \Psi_{\xi}^{i}(x)^\dagger \,
\Psi_{\xi'}^{j}(x) \CR
&=& \frac{1}{2 \pi} \int_{T^{2p}} d^{2p} x' \, 
\sum_{l_1,\cdots,l_{2p} \in \bZ } \,
\exp \left(
i \frac{(x'^\mu+2 \pi R_\mu l_\mu)(\xi^\mu-\xi'^\mu)}{2 \pi \alpha'}
\right)
\,\psi_{\xi}^{i}(x)^\dagger \,
\psi_{\xi'}^{j}(x) \CR
&=& \int_{T^{2p}} d^{2p} x' \, \psi_{\xi}^{i}(x')^\dagger \,
\psi_{\xi'}^{j}(x') 2 \pi \delta (\xi-\xi'),
\eeqa
where $x^\nu= x'^\nu+2 \pi R_\nu l_\nu $
and we have used 
$\psi_{\xi}^{i}(x)=\Omega^{-1}_\nu \psi_{\xi}^{i}(x')$
which implies $\psi_{\xi}^{i}(x)^\dagger \,
\psi_{\xi'}^{j}(x)= \psi_{\xi}^{i}(x')^\dagger \,
\psi_{\xi'}^{j}(x')$.

Now we can evaluate the coordinate $\bar{X}^\mu$ of 
$m$ D0-branes corresponding 
to the remaining $m$ zero modes:
\beqa
\left(\bar{X}^\mu  \right)^{i, \xi}_{\;\;\;\; j, \xi'}
&=&
 \frac{1}{2 \pi}
\int_{\bR^{2p}} d^{2p} x \, \Psi_{\xi}^{i}(x)^\dagger \,
x^\mu \, \Psi_{\xi'}^{j}(x) \CR
&=&
\alpha' \int_{\bR^{2p}} d^{2p} x \, 
\Psi_{\xi}^{i}(x)^\dagger \,
\left( -i \frac{\de}{\de \xi'^\mu} 
\Psi_{\xi'}^{j}(x)   
+i e^{i \frac{1}{2 \pi \alpha'} \xi'_\mu  x^\mu  }
\frac{\de}{\de \xi'^\mu} \psi_{\xi'}^{j}(x)   
\right)
\CR
&=&
2 \pi \alpha' \left(
i \delta_{ij} \frac{\de}{\de \xi^\mu}
+\left( \tilde{A}_\mu (\xi) \right)^{i}_{\;\; j} 
\right) \delta (\xi-\xi'),
\label{dualX}
\eeqa
where
\beq
\left( \tilde{A}_\mu (\xi) \right)^{i}_{\;\; j}
= i \int_{T^{2p}} d^{2p} x' \, \psi_{\xi}^{i}(x')^\dagger \,
\frac{\de}{\de \xi^\mu} \psi_{\xi}^{j}(x').
\label{dualA}
\eeq
This means that 
\beq
\bra{\xi,i} \bar{X}^\mu 
= 2 \pi \alpha' \left(
i \delta_{ij} \frac{\de}{\de \xi^\mu}
+\left( \tilde{A}_\mu (\xi) \right)^{i}_{\;\; j} 
\right) \bra{\xi,j},
\eeq
thus $ \bar{X}^\mu 
= 2 \pi \alpha' \left(
i \delta_{ij} \frac{\de}{\de \xi^\mu}
+\left( \tilde{A}_\mu (\xi) \right)^{i}_{\;\; j} 
\right)$ in this basis.
Moreover, from
\beq
U_\nu  \Psi_{\xi}^{i}(x)=e^{i \frac{\xi^\nu}{L_\nu}}  
\Psi_{\xi}^{i}(x),
\eeq
for $U_\nu=  \Omega_\nu \, e^{2 \pi i \hat p_\nu R_\nu}$,
we find an equivalence between 
the $N$ D$(2p)$-branes on the $T^{2p}$ with the gauge field
$A_\mu(x)$
and the $m$ D0-branes on the {\it same} $T^{2p}$ with
the coordinates $\bar{X^\mu}$ (\ref{dualX}).
From the relation (\ref{rep1}),
the T-dual of the latter D0-branes is 
$m$ D$(2p)$-branes on the dual torus $\tilde{T}^{2p}$ 
with the gauge field $\tilde{A}_\mu(\xi)$.
Note that 
\beq
m= \int_{T^{2p}} \Tr e^{\frac{F}{2 \pi}}, \;\;\;
N= \int_{\tilde{T}^{2p}} \Tr e^{\frac{\tilde{F}}{2 \pi}}, \;\;\;
\eeq
are followed from the index theorem.

Therefore, 
we find an equivalence 
between the $N$ D$(2p)$-branes on the $T^{2p}$ with the gauge field
$A_\mu(x)$ and
the $m$ D$(2p)$-branes on the dual torus $\tilde{T}^{2p}$ 
with the gauge field $\tilde{A}_\mu(\xi)$ given by (\ref{dualA})
using the Dirac zero modes (\ref{dpsi}).

The transition function for  $\tilde{A}_\mu(\xi)$
is given by
\beq
\left( \tilde{\Omega}_\nu(\xi)  \right)^{i}_{\;\; j}
= \int_{\bR^{2p}} d^{2p} x \, \Psi_{\xi}^{i} (x)^\dagger \,
\Psi_{\xi'}^{j}(x) 
=
\int_{T^{2p}} d^{2p} x' \, \psi_{\xi}^{i}(x')^\dagger \,
{\psi^{(\nu)}}_{\xi}^{j}(x'),
\eeq
where 
\beq
\xi'_\mu=\xi_\mu+ 2 \pi \delta_{\mu \nu} L_\nu,
\eeq
and
\beq
{\psi^{(\nu)}}_{\xi}^{j}(x)=
e^{\frac{L_\nu x^\nu}{\alpha'}}
{\psi}_{\xi'}^{j}(x),
\eeq
which satisfies
$\Dslash_\xi \; {\psi^{(\nu)}}_{\xi}^{j}(x)=0 $,
thus a linear combinations of $ \psi_{\xi}^i (x) $.

If we take $p=2$ and $A_\mu(x)$ is anti-self dual,
the formula (\ref{dualA}) is indeed 
the Nahm transformation of \cite{BrBa, Sc},
which is a generalization of the formula 
given in \cite{CoGo} for the ADHM case.
We note that 
the Nahm transformation can be viewed as 
a combination of the 
two different equivalences:
(1) the T-dual and 
(2) the equivalence between the $N$ D4-brane with $A_\mu$ and 
the $m$ D0-branes with $\bar{X^\mu}$.

\subsection{T-dual of the $\dz$-brane pairs}

Let us take the T-dual of 
the D$(2p)$-brane solution on the torus, (\ref{solt}).
Now we assume that the bundle on the D$(2p)$-brane is trivial
and $A_\mu(x)=\zeta_\mu/(2 \pi \alpha')$, where $\zeta_\mu$ is a constant.
In this case, $U_\nu$ is just a translation operator.
As we have seen, the Hilbert space of the Chan-Paton index is spanned by
the spinors on the $\bR^{2p}$ and 
any spinor $\Psi(x)$ on $\bR^{2p}$ can be
written as
\beqa
\Psi(x) &=& \int_0^{\frac{2 \pi \alpha'}{R_1}} d \xi_1 
\int_0^{\frac{2 \pi \alpha'}{R_2}} d \xi_2 \cdots
\int_0^{\frac{2 \pi \alpha'}{R_{2p}}} d \xi_{2p} 
\sum_{n_{\mu} \in \bZ} \;
e^{i \frac{1}{2 \pi \alpha'} \left( \xi_\mu + \frac{2 \pi
\alpha'}{R_\mu} n_\mu \right) x^\mu  }
\psi(\xi,n) \CR
 &=& \int_0^{\frac{2 \pi \alpha'}{R_1}} d \xi_1 
\int_0^{\frac{2 \pi \alpha'}{R_2}} d \xi_2 \cdots
\int_0^{\frac{2 \pi \alpha'}{R_{2p}}} d \xi_{2p} 
e^{i \frac{1}{2 \pi \alpha'} \xi_\mu  x^\mu  } \;
\psi_\xi(x),
\eeqa
which is just a Fourier transformation with the momentum
$p_\mu= \frac{1}{2 \pi \alpha'} \left( \xi_\mu + \frac{2 \pi
\alpha'}{R_\mu} n_\mu \right)$.
Here we defined $\psi_\xi(x)= \sum_{n_{\mu} \in \bZ} \;
e^{i \frac{1}{2 \pi \alpha'} \frac{2 \pi \alpha'}{R_\mu} 
n_\mu  x^\mu  } \psi(\xi,n)$
which is periodic, namely, $U_\nu \psi_\xi(x)=\psi_\xi(x)$
and $ \psi(\xi,n)$ is a constant spinor.
Then, 
\beq
\Psi_{\xi,n}(x)
= e^{i \frac{1}{2 \pi \alpha'} \left( \xi_\mu + \frac{2 \pi
\alpha'}{R_\mu} n_\mu \right) x^\mu  } \psi,
\eeq
is a basis of the Hilbert space labeled by $\{ \xi, n \}$
and the spinor index of a constant spinor $\psi$.
We note that $\Psi_{\xi,n}(x)$ is an eigen state of $U_\nu$, 
\beq
U_\nu \Psi_{\xi,n}(x) =   
e^{i \frac{R_\nu}{\alpha'} \xi_\nu  }  \Psi_{\xi,n}(x)
\eeq
and also an eigen state of the tachyon 
$T=\lim_{u \ra \infty}u  \; \Dslash$,
\beq
\Dslash \; \Psi_{\xi,n}(x)
=\frac{1}{2 \pi \alpha'} 
\Gamma^\mu \left(  \xi_\mu + \frac{2 \pi
\alpha'}{R_\mu} n_\mu  -\zeta_\mu \right) \;
\Psi_{\xi,n}(x).
\eeq
In this basis, $X^\mu=\hat x^\mu$ is represented 
as $2 \pi \alpha' i \frac{\de }{\de \xi^\mu}$. 
This means that 
the gauge fields of
the D$(2p)$-branes and anti-D$(2p)$-branes
in the T-dual picture vanish.

Now we expect that 
the T-dual of the tachyon will
be given by the tachyon in the above basis of the Chan-Paton
bundle
since we regard the torus as the orbifold 
of $\bR^{2p}$.
Therefore, in the T-dual picture,
this system is equivalent to 
infinitely many pairs of D$(2p)$-brane and anti-D$(2p)$-brane,
labeled by $\{ n_\mu  \} \in \bZ^{2p}$, on the dual torus $\tilde{T}^{2p}$.
The tachyon condensation is given by
\beq
\tilde{T}(\xi)= \tilde{u} \Gamma^\mu \left(\xi_\mu-\zeta_\mu+2 \pi L_\mu n_\mu 
\right),
\label{tdualconf}
\eeq
which is diagonal in $n_\mu$
and $\tilde{A}_\mu=0$.
Here we defined $\tilde{u}=\frac{u}{ 2 \pi \alpha'}$.
We interpreted that the $\{ \xi_\mu \}$ parameterize
the world volume of the pairs of 
$\mbox{D}2{\rm -}\overline{\mbox{D}2}$-branes,
on the other hand, 
$\{ n_\mu\}$ are the Chan-Paton indices.
$\zeta_\mu$ is the location of the 
a solitonic D$0$-branes on the dual torus $\tilde{T}^{2p}$.

Since $\Psi_{\xi' ,n}=\Psi_{\xi,n'}$,
where $\xi'_\mu=\xi_\mu +2 \pi \delta_{\mu \nu} L_\nu$ and 
$n'_\mu=n_\mu+\delta_{\mu \nu}$,
the transition function of the infinitely many 
pairs of D$(2p)$-branes and anti-D$(2p)$-branes
in this T-dual picture, 
is given by
\beq
\tilde{\Omega}_\nu=U_{n_\mu \rightarrow  n_\mu+\delta_{\mu, \nu} },
\eeq
where $U_{n_\mu \rightarrow  n_\mu+\delta_{\mu, \nu} }$ 
is the unitary operator
which maps $\Psi_{\xi,n}$ to $\Psi_{\xi,n'}$.
Thus the tachyon (\ref{tdualconf}) is a consistent configuration 
on the dual torus although it is not periodic.\footnote{
We thank Koji Hashimoto for suggesting this solution.}

We note that the configuration (\ref{diracop}) of
the $\dz$ pairs is more convenient than
its T-dual configuration (\ref{tdualconf}) 
of D$(2p)$-anti-D$(2p)$ pairs, especially, 
for a configuration with a non-trivial $A_\mu(x)$.
For a non-trivial $A_\mu(x)$,
from a spinor $\psi_{\xi,n} (x)$ 
stisfying
\beq
U_\nu \psi_{\xi,n}(x) =  \psi_{\xi,n}(x),
\;\;\;\; 
\Dslash_\xi \psi_{\xi,n}(x)=E_{\xi,n} \psi_{\xi,n}(x),
\eeq
a basis is given by 
\beq
\Psi_{\xi,n} (x) 
= e^{i \frac{1}{2 \pi \alpha'} \xi_\mu  x^\mu  }
\psi_{\xi,n} (x),
\label{decomp2}
\eeq
where $\Psi_{\xi,n} (x) $ 
is an eigen state of $U_\nu$
and $\Dslash $. 
Then, the tachyon configuration 
of D$(2p)$-anti-D$(2p)$ pairs on $\tilde{T^{2p}}$
is implicitly given by
\beq
\tilde{T}(\xi) = u \; E_{\xi,n}
\eeq
in this basis.

Let us take the large radius limit of the torus or 
the T-dual torus.
If we take the $L_\mu \rightarrow \infty$,
then only the pair of D$(2p)$-brane and anti-D$(2p)$-brane
with $n_\mu=0$ will remain and
the configuration (\ref{tdualconf}) becomes
\beq
\tilde{T}(\xi)= \tilde{u} \gamma^\mu \left(\xi_\mu-\zeta_\mu 
\right),
\label{decent1}
\eeq
which is just 
the Atiyah-Bott-Shapiro solution \cite{KMM2, KrLa, TaTeUe} for the 
decent relation.
On the other hand,
if we take the $R_\mu \rightarrow \infty$,
the original  
infinitely many $\dz$-brane pairs on $T^{r}$
become those on $\bR^{r}$
and the solution (\ref{solt}) is same 
as the solution for the ascent relation found in \cite{Te, AsSuTe1}.
Thus we can say that 
on the torus the two solutions
for the decent relation (\ref{decent1}) 
and the ascent relation (\ref{solt})
are T-dual each other.

Finally, we will comment on 
the classification 
of the D-branes by the K-theory.
The configuration (\ref{decent1}) for the 
decent relation is related to the K-theory
(using the infinitely many 
$\mbox{D}9{\rm -}\overline{\mbox{D}9}$ pairs \cite{HaMo, Wi2}).
On the other hand,
(\ref{solt}) represnts 
the (analytic) K-homology class
in \cite{AsSuTe1}.
Therefore, 
we expext that 
the T-dual maps the K-theory
to the analytic K-homology.
However, 
the winding modes are neglected to obtain the analytic K-homology
by assuming the size of the compactified manifold is very large
in \cite{AsSuTe1}
although the winding modes are important for 
the T-dual picture.
The duality of the KK-theory discussed
in \cite{Mathai} will be important to study the 
role of the widing modes.
It would be interesting to investigate the relation to it
further.

\section{Concluding remarks}
\label{sec:4}

In this paper, we found 
an exact solution, with a nonzero net D-brane charge,
of the tachyon condesation
in the boundary string field theory 
of brane-anti-brane pairs on torus.
The Nahm-transformation of the instantons 
was derived from this tachyon condensation.
We also found 
the T-dual configutration of this.

There are several interesting future directions. 
Since our method is not restricted to the instanton
(i.e. $p=2$) case,
it will be interesting
to study the Nahm transformation for D0-D8 or D0-D6 cases.
Morevoer, the BPS properties are not 
(explicitly) assumed in this paper.
Therefore, the non-BPS cases, for exampole models dicussed in 
\cite{Ta2} \cite{HaKa}, \cite{Wi}
are also covered in thie paper.
To extend our result to other orbifolds,
like ALE spaces, are also interesting.

In this paper, 
we do not explicitly use 
the boundary state formalism
though we believe the exact solutions in the 
boundary string field theory can be mapped to
the boudnary state.
(The marginal deformation case \cite{MaSe} was 
studied in \cite{NaTaUe}.)
It would be desired to do it explicitly.

\acknowledgments 
S.~T.~ is grateful to K.~Hashimoto and S.~Sugimoto for
useful comments and discussions. 
S.~T.~is partly supported by
the Japan Ministry of Education, Culture, Sports, Science and
Technology. 


\newcommand{\J}[4]{{\sl #1} {\bf #2} (#3) #4}
\newcommand{\andJ}[3]{{\bf #1} (#2) #3}
\newcommand{\AP}{Ann.\ Phys.\ (N.Y.)}
\newcommand{\MPL}{Mod.\ Phys.\ Lett.}
\newcommand{\NP}{Nucl.\ Phys.}
\newcommand{\PL}{Phys.\ Lett.}
\newcommand{\PR}{ Phys.\ Rev.}
\newcommand{\PRL}{Phys.\ Rev.\ Lett.}
\newcommand{\PTP}{Prog.\ Theor.\ Phys.}
\newcommand{\hep}[1]{{\tt hep-th/{#1}}}

\end{document}